\documentclass[10pt,conference,compsocconf,final]{IEEEtran}
\usepackage{times}

\usepackage{caption}
\ifCLASSINFOpdf
\else
\fi
\usepackage{color}
\usepackage{soul}

\begin{document}
\pagenumbering{gobble}
\title{\textbf{\Large Overview on Security Approaches in Intelligent Transportation Systems\\} \large Searching for hybrid trust establishment solutions for VANETs\\[0.2ex]}
%[-1.5ex]
%\author{\IEEEauthorblockN{~\\[-0.4ex]\large Christoph Ponikwar\\[0.3ex]\normalsize}
%\IEEEauthorblockA{MuSe - Munich IT Security Research Group\\Department of Computer Science and Mathematics\\Munich University of Applied Sciences (MUAS)\\
%Lothstra\ss e 64, 80335 Munich, Germany\\
%Email: {\tt christoph.ponikwar@hm.edu}\\
%Telephone: +49 (0) 89 1265--3792\\
%Fax: +49 (0) 89 1265--3780}
%\and
%\IEEEauthorblockN{~\\[-0.4ex]\large Hans-Joachim Hof\\[0.3ex]\normalsize}
%\IEEEauthorblockA{MuSe - Munich IT Security Research Group\\Department of Computer Science and Mathematics\\Munich University of Applied Sciences (MUAS)\\
%Lothstra\ss e 64, 80335 Munich, Germany\\
%Email: {\tt hof@hm.edu}\\
%Telephone: +49 (0) 89 1265--3752\\
%Fax: +49 (0) 89 1265--3780}}

\author{%
\IEEEauthorblockA{~\\[-0.4ex]\large Christoph Ponikwar, Hans-Joachim Hof\\[0.3ex]\normalsize}
\IEEEauthorblockA{MuSe - Munich IT Security Research Group\\Department of Computer Science and Mathematics\\Munich University of Applied Sciences (MUAS), Germany}
e-mail: {\tt christoph.ponikwar@hm.edu}, {\tt hof@hm.edu} %
}


%Telephone: (800) 555--1212, Fax: (888) 555--1212}


\maketitle

\begin{abstract}
%\boldmath
%In the recent years further advancements Cyber-Physical Systems (CPSs) in form of intelligent transportation systems, regardless whether autonomous driving or assisted driving, have been made. 
Major standardization bodies developed and designed systems that should be used in vehicular ad-hoc networks. The Institute of Electrical and Electronics Engineers (IEEE) in America designed the wireless access in vehicular environments (WAVE) system. The European Telecommunications Standards Institute (ETSI) did come up with the ``ITS-G5'' system. Those Vehicular Ad-hoc Networks (VANETs) are the basis for Intelligent Transportation Systems (ITSs). They aim to efficiently communicate and provide benefits to people, ranging from improved safety to convenience. But different design and architectural choices lead to different network properties, especially security properties that are fundamentally depending on the networks architecture. To be able to compare different security architectures, different proposed approaches need to be discussed. One problem in current research is the missing focus on different approaches for trust establishment in VANETs. Therefore, this paper surveys different security issues and solutions in VANETs and we furthermore categorize these solutions into three basic trust defining architectures: \textit{centralized}, \textit{decentralized} and \textit{hybrid}. These categories represent how trust is build in a system, i.e., in a centralized, decentralized way or even by combining both opposing approaches to a hybrid solution, which aims to inherit the benefits of both worlds. This survey defines those categories and finds that hybrid approaches are underrepresented in current research efforts. 
\end{abstract}



% \fi
\IEEEpeerreviewmaketitle

\section{Introduction}
\label{introduction}
%\textbf{\color{red}FIXME Big Picture und Details nur wenn notwendig!
%FIXME Introduction Problemstellung des Papers ist nicht klar genug, welches Problem will das Paper lösen? Related Work wahrscheinlich am besten nach der Einleitung um zu begründen, warum es Ihr Paper braucht}

This paper surveys different security architecture techniques for VANETs used in ITSs. Security plays a significant role in modern Cyber-Physical Systems (CPSs), which include intelligent transport systems. Trust establishment describes how trust is formed, which in return defines the fundamental security architecture of a system. The future of ITS is a networked one where vehicles and infrastructure do communicate to make traffic more efficient and safer. As vehicles are inherently mobile and self containing, the only way to communicate on the move is via wireless technology. Wireless has proven over time that getting security right in wireless technology is hard. An example of security done wrong is the utterly broken WEP technology, which is an acronym for ``Wire Equivalent Privacy'', but it was never able to fulfill that promise. Not only do security issues in regards to authentication exist but furthermore there are also security issues, like denial of service, replay or spoofing attacks, which vary in severity and easy of exploitability. But overall security in wireless technologies is boiling down to how trust is established and which methods and algorithms are used to secure the trust establishment.

One centralized approach for conveying trust is the mode of operation that is used in telecommunication standards all over the world, i.e., Global System for Mobile Communications (GSM), Code division multiple access (CDMA), Universal Mobile Telecommunications System (UMTS) or Long-Term Evolution (LTE). That mode of operation is building trust based on a shared symmetric secret. Until recently the weaknesses were discussed, like the important dependence of secrecy of this shared secret, but arguments were often discarded because of the needed effort to steal the secret key of each customer and the high security approach network operators supposedly are taking to secure those secrets. The treasure trove of leaked information by former National Security Agency (NSA) contractor Edward Snowden, showed the security community again once more how bad our assumptions were in this regard. As documents provided by the online publication THE INTERCEPT \cite{great-sim-heist-2015} show that American (NSA) and British spy agencies Government Communications Headquarters (GCHQ) managed to steal those important secrets directly from the manufacturer, in this case Gemalto. The theft means, on a technical level, that authenticity and confidentiality of a communication, supposedly secured by those stolen secrets, is compromised. The methods supposedly used to execute that theft raise many questions but this discussion might fit better in a legal or social publication and should not be discussed in this paper. Another rather recently uncovered attack on several banks (ca. 100 banks) around the world (ca. 30 countries) where a so called ``Carnbanak cybergang'' have stolen an estimated amount of 1 billion United States dollar (USD). While the used malware does not appear to be of very high sophistication, only the weakest trust relationship in computing, between a human and a machine was exploited via spear phishing, the orchestration, endurance and the targeted approach of the attackers where extremely remarkable, as reported by Kapersky Labs \cite{kaspersky-lab-hq-carbanak-apt-eng.pdf-2015}. 

The underestimation, to what length state actors would go to achieve informational advantage in combination with how persistent and patient criminal actors are becoming, previously only attributed to state actors, goes to show how wrong and weak our current assumptions on cybersecurity have been. All central approaches bear the risk of being exploited by targeted attacks on the central trust anchors. And this is why we urge every researcher to reevaluate their assumptions and seek for alternative designs. Both currently developed major standards, IEEE WAVE \cite{_1609.0-2013_2014} or ETSI ITS-G5 \cite{etsi_tr_102_962_etsi_2012}, favor a centralized security architecture, with trust rooting in central authorities, which represent a high value target for attackers to exploit.

Decentralized approaches could be such an alternative, by making such attacks much more risky and costly due to the distribution of trust relations. Distributed trust relations have their own issues, like performance or new attack opportunities not present in centralized architectures. Therefore, the combination of both architectural approaches, in this paper called hybrid approaches, could pose a overall security improvement, especially in the current environment with increasing proliferation of attack and exploitation techniques accessible to criminals and state actors alike. Attacking seams to be easier than defending, this is why we argue a easy to defend security architecture is paramount for any information system or network nowadays, especially in the field of ITSs, which are in focus of this paper. 

As stated previously trust establishment can be achieved via a centralized way e.g., a Public Key Infrastructure (PKI), decentralized e.g., a Web of Trust (WoT) or by using a hybrid approach, which tries to combine the benefits of both approaches. Security issues in mobile ad-hoc networks are used to find solution for them and then categorizing those solutions into their general security architecture. We limit ourselves to some of the following major issues in ITS mentioned by various researches like Hubaux et al.\cite{Hubaux2004}, Lin et al.\cite{Lin2008} or in an already summarized from by Yang \cite{Yang2013}.
%
%\textbf{\color{red}FIXME Auswahl erscheint willkürlich...}

\begin{itemize}
\item Impersonating by false, stole identities, Message spoofing or replay attacks
\item Tampering with data in-transit
\item Send false feedback to silence other vehicles
\item Sinkhole attack via false routing information to effectively execute Man-in-the-Middle (MitM) attacks
\item Sybil attack by creating virtual sock puppet identities to manipulate voting procedures to the attackers benefit
\item An eclipse attack is similar to an sybil attack it specifically tries to split a network by using means of a malicious group of nodes
\item Manipulating the network topology and disturbing node by connecting far away segments via a hidden tunnel (wormhole attack)
\item Privacy violation caused by continues communication
\item Denial of Service (DoS) by jamming signals or overloading specific nodes
\end{itemize} 

In \cite{Agrawal}, Agrawal et al. present a short overview of different security issues and solutions with their objectives and draw backs.
Mishra et al. \cite{Mishra2011} display a wide array of research effort in regards to security issues and solutions, which they think are important. A detailed introduction into VANETs is given by Raya et al. \cite{Raya2007a} they furthermore expand on, security issues and solutions in VANETs. Zhang\cite{Zhang2011} categorizes trust management for VANETs in three models: \textit{Entity-oriented Trust Model}, \textit{Data-oriented Trust Model} and \textit{Combined Trust Model}. We differentiate various approaches to security issues in VANETs into three categories: \textit{Centralized}, \textit{Decentralized} and \textit{Hybrid} as we think these categories describe the way trust is build better. 

We define those categories in detail in the following Section \ref{analysis}. Thereafter in separate sections we describe eight different security issues, already defined by Yang \cite{Yang2013}. Each of these section contains solutions to its security issues, which are categorized according to our definition into: \textit{Centralized}, \textit{Decentralized} and \textit{Hybrid} solutions. A summary of this paper is provided in the last Section \ref{conclusion}.

\section{Analysis}
\label{analysis}
One of the main issues in ad-hoc-communication is trust. It is a basic problem in security to establish so-called trust anchors. Several models for trust management exist. The surveyed approaches with their assigned categories are listed in the summary table \ref{table_analysis_overview}. 

\textbf{Centralized}: A central trust model may for example be implemented by a PKI. A PKI consists of one or more Certificate Authorities (CAs) that issue certificates to the participants of the system. The issue of certification may be delegated to Sub-CAs, resulting in a hierarchy of CAs. A certificate of a participant is considered to be legitimate, if it is possible to find a certification path from the certificate to a known and trusted CA. Several (yet unknown) Sub-CAs may be on the certificate path. Per se, all legitimate certificates are considered trusted. Hence, all the known and trusted certificates represent trust anchors for one participant. Using a PKI simplifies trust establishment to secure setup of trust anchors on an instance of the system. 

\textbf{Decentralized}: A decentralized trust model may for example be a so WoT. In a WoT scenario, each participant of the whole network is also a CA and may express trust in a certificate of another participant. Each participant keeps a list of other participants of the system that are trusted and another list of participants that are trusted to express trust in other participants. As with the PKI, establishing trust in an unknown participant requires to build a trust path between the unknown participant and oneself. However, as no hierarchy exists, finding such a path is a hard task. Another approach to trust establishment are reputation models. No certificates are issued but the behavior of participants is monitored and trust values are assigned based on different attributes like former or expected behavior. Participants may exchange trust values of each other.

\textbf{Hybrid}: A hybrid trust model is one that makes use for example of a distributed PKI, which assigns identities to participants, mainly for liability reasons. This trust path is only used in case of an accident or when certain conditions are met. But the operational trust between participants is realized via a reputation system and only if enough evidence of bad or malicious behavior was recorded, the PKI infrastructure would step in to permanently revoke or destroy the cryptographic material of the offending node. Hybrid solutions are trying to combine both central and decentral approaches, to get the benefits of both approaches, like somewhat independence of central infrastructure or better privacy features.

%\begin{table}[!t]
\begin{table}[b]
\renewcommand{\arraystretch}{1.3}
% if using array.sty, it might be a good idea to tweak the value of \extrarowheight as needed to properly center the text within the cells
\caption{ANALYSIS OVERVIEW}
\label{table_analysis_overview}
\centering
\begin{tabular}{|l||c|c|c|c|}
\hline
Security Issues & Centralized & Decentralized & Hybrid & Ref.\\
\hline \hline
Impersonation & \cite{Sun2007a},\cite{Hubaux2004},\cite{Raya2007} & \cite{Golle2004},\cite{Balfanz2002} & \cite{Wasef2009a} & \ref{impersonation}\\
Data Tampering & \cite{Li2008},\cite{Sun2007a},\cite{Raya2007a} & \cite{Golle2004} & \cite{Zhang2008} & \ref{data-tampering}\\
Routing Attacks & \cite{Eichler2004},\cite{Lu2010},\cite{Zhong2003} & \cite{Huang2003},\cite{Buchegger2002} & \cite{Raya2007} & \ref{routing-attacks}\\
Sybil Attacks & \cite{Piro2006} & \cite{Xiao2006},\cite{Patcha2007} & \cite{Park2009} & \ref{sybil-attack}\\
Eclipse Attacks & \cite{Wasef2009} & \cite{Xiao2006} & - & \ref{eclipse-attack}\\
Wormhole Attacks & - & \cite{Safi2009},\cite{Eichler2004} & - & \ref{wormhole-attack}\\
Denial of Service & \cite{Raya2007} & \cite{Hamieh2009} & - & \ref{denial-of-service}\\
Privacy Violation & \cite{Sun2007a},\cite{Choi2009},\cite{Wasef2009a},\cite{Wasef2009},\cite{Wasef2010} & \cite{Li2008} & \cite{Bechler2004},\cite{Zhou1999} & \ref{privacy-violation}\\
\hline
\end{tabular}
\end{table}

\subsection{Impersonation}
\label{impersonation}
Defending against replay or whole message spoofing attacks is usually done at a protocol level. If used communication protocols do not defend against those attacks a communication system, regardless its architecture will be hard to secure. Communication systems usually use some kind of identities to distinguish between different participants. Those identities usually need to be protected against impersonation to sustain distinguishability.

\textbf{Centralized}: In case strong identities are needed like in a system utilizing identity based cryptography \cite{Boneh2001} Sun et al. \cite{Sun2007a} propose storing identities in a tamper proof hardware to prevent identity theft. So do Hubaux et al. \cite{Hubaux2004} they store their form of identity, called electronic license plate, in an event date recorder (EDR), similar to a black box in an aircraft. The EDR in return itself should be ``protected [...] physically''\cite{Hubaux2004}. Similarly Raya et al. \cite{Raya2007} are using a ``trusted component'' in their protocols to store and protect identity data against theft. 

\textbf{Decentralized}: Every participant in a VANET should have its own model of its vicinity and validate every piece of data received, according to Golle et al. \cite{Golle2004}. They authenticated communication via public/private key pairs but they are self generated by each node and should be refreshed constantly \cite{Golle2004}. Additionally, they propose using ``location-limited channels''\cite{Balfanz2002} to distinguish nodes. As an example of a ``location-limited channel''\cite{Balfanz2002} is the use of infrared signaling given by Golle et al. \cite{Golle2004}. 

\textbf{Hybrid}: In an approach called ``Efficient Decentralized Revocation Protocol''\cite{Wasef2009a} (EDR) Wasef et al. propose a way to revoke trust in identities based on ``probabilistic random key distribution technique and a novel pairing-based threshold scheme''\cite{Wasef2009a}. It uses PKI but the revocation process is decentralized and facilitated by voting.

\subsection{Data Tampering}
\label{data-tampering}
Depending on how nodes are communicating in a VANET, whether it is single hop or multi hop communication, different opportunities arise for data tampering or manipulation. If transmitted data is not integrity protected, any intermediate system or bystander could change the information for its own benefit.

\textbf{Centralized:} One of the more complete approaches was proposed by Li et al. \cite{Li2008}. They based their scheme also on identity based cryptography \cite{Boneh2001}. But they extended it with blind signatures and one-way hash chains to provide mutual authentication, confidentiality and integrity, while preserving privacy. This approach is  similar to that of Sun et al. \cite{Sun2007a}, which also is based on identity cryptography and aims to deliver on the same security requirements \cite{Raya2007a}.

\textbf{Decentralized}: Using a reputation system in conjunction with collecting and querying for additional data, to verify and attest trustworthiness of information is proposed by Golle et al. \cite{Golle2004}. Every node builds up his own model of the network around him and validates data against it.

\textbf{Hybrid}: In \cite{Zhang2008}, Zhang et al. present a scheme called ``RAISE'' a Roadside Unit(RSU)-adied message authentication scheme, which uses keyed-hash message authentication where the secret key is known by the RSU, which in return can therefore attest that the message is authentic. The proposed scheme is compatible with traditional PKI-based systems, further more it makes use of PKI as a fallback mechanism.

\subsection{Routing Attacks}
\label{routing-attacks}

To prevent congestion in ad-hoc wireless environments nodes are listening to its neighbors and if a neighbor is better suited to forward messages it stops rebroadcasting messages. If an attacker could convince a node that he is better positioned, the attacker can silence other nodes. Which would make them effectively disappear from the VANET, so called silencing attacks.
Also a vehicular ad-hoc network where bandwidth is limited, and far reaching connections to central systems needed to be routed through long range wireless communication technology like LTE or UMTS. Those communication technologies are expensive to use compared to a node posing as a high speed uplink or gateway reachable via ad-hoc communication, called sinkholing attack. This enables MitM attacks, where a malicious gateway can intercept or even alter the sent and received messages.

\textbf{Centralized}: One of the first secure routing protocols for VANETs were proposed by Eichler et al. \cite{Eichler2004} called  AODV-SEC based on ``Ad-hoc On-demand Distance Vector'' (AODV). Lu et al. designed the ``social-based privacy-preserving packet forwarding'' \cite{Lu2010} (SPRING) to be resistant against black holing attacks by utilizing road side infrastructure. Relying on PKI for strong identities but giving incentives, based on game theory, to nodes taking part in a mobile ad-hoc network, was proposed by Zhong et al. \cite{Zhong2003}. Sprite, ``a simple, cheat-proof, credit-based system for mobile ad-hoc networks'' \cite{Zhong2003} also needs a central Credit Clearance Service (CCS) to function. 

\textbf{Decentralized}: In \cite{Huang2003}, Huang et al. propose a cluster based intrusion detection system to detect various attacks, among them sinkholing or blackholing. Their approach is focused on detection of those attacks and mitigation is left for the network to handle. The CONFIDANT protocol by Buchegger et al. \cite{Buchegger2002} consist out of four entities present in each node: Monitor, Reputation System, Path Manager and Trust Manager. The Trust Manager collects events via the Monitor and uses the Reputation System to evaluate the events and the result of the evaluation are used by the Path Manager to adjust the rounting, to mitigate attacks like sinkholing. 

\textbf{Hybrid}: To detect and respectively mitigate misbehaving nodes Raya et al. \cite{Raya2007} propose two methods ``Misbehavior Detection System (MDS)'' and ``Local Eviction of Attackers by Voting Evaluators (LEAVE)''. When detecting a misbehaving node, LEAVE is used to degrade the attackers trust until a central certificate authority revokes its certificates. LEAVE is resilient to interference as long as colluding attackers are a minority.

\subsection{Sybil Attacks}
\label{sybil-attack}
When protocols with voting procedures are used or if some kind of collaboration between nodes for making collective group decisions is needed, then a so called sybil attack could be used to influence protocols or decisions. This is done by creating sock puppets that the attacker controls to act on behave of him. In an vehicular environment, if an attacker would like to push the envelope, he and his sock puppets could simulate braking or congestion, and then tricking the victims into believing him. Protocols like the previously mentioned LEAVE Protocol by Raya et al. \cite{Raya2007} have a certain threshold to, which they are resilient against a sybil attack. The important factor is the size of the sock puppet group in comparison to the amount of honest nodes.

\textbf{Centralized}: An easy protection against sybil attacks is the use of centrally enforced and distributed strong identities. Identities are created by a central authority and handed down to the nodes prior to their deployment as stated by Piro et al. \cite{Piro2006}. This process could be upfront or part of a VANET joining protocol. Either way a central entity knows to whom it has handed a specific identity. With autonomous vehicles at the horizon it may be even more compelling or tempting to use the vehicle identification number as such an id. This approach has many privacy implications, like unique traceable identities, or the central data storage would be a high value target for theft or intrusion.  

\textbf{Decentralized}: In \cite{Xiao2006}, Xiao et al. draft a technique called ``Basic Signal-Strength-Based Position Verification'' \cite{Xiao2006}, which is used to verify the position by a claimer based on the signal strength. This technique is then used after collecting beacon messages to decide based on probability if there is a sybil node nearby and if so a statistic model is used to attribute the sybil nodes to one originating vehicle. Park et al. are using a timestamp based approach to detect sybil attackers \cite{Patcha2007}.

\textbf{Hybrid}: An approach using timestamp series and RSUs issuing certificates was proposed by Park et al. \cite{Park2009}. The RSUs themselves have public private key pairs and a certificate from a central certificate authority. All vehicles must have the public key of the certificate authority pre-installed. Additional vehicles generate their own pair of keys. Similar timestamps series are identified as a sybil attack. To protect against sybil attack each data message must contain current timestamp certificate, RSU certificate, signed data and of course the data itself. If any inconsistencies occur the packets should be dropped. As the authors suggested by themselves \cite{Park2009} this approach is not suited for high traffic and urban scenarios, due to the spatial and temporal difference assumption falling apart.     

\subsection{Eclipse Attacks}
\label{eclipse-attack}
An eclipse attack utilizes compromised neighbors to influence group decisions. It is also useful when the separation of nodes from other nodes weakens the whole network segment, by degrading the trust in the honest group while improving its own standing in the network. This approach usually eases and strengthens other attacks like DoS \ref{denial-of-service}.

\textbf{Centralized}: Quick and efficient removal of identified malicious nodes is key in protecting against eclipse attacks. Therefore Wasef et al. \cite{Wasef2009} proposed, based on a PKI system, not only a novel message authentication approach but also a quick certificate revocations approach to evict the trustworthiness of misbehaving nodes.

\textbf{Decentralized}: Some of the methods used to defend against sybil attacks also could be used to defend against eclipse attacks especially Xiao et al. \cite{Xiao2006} are trying to suppress sybil attacks in conjunction with opposite traffic flow and their ability to proof that they came from an upstream source.

\textbf{Hybrid}: \textbf{-} No hybrid approaches were found in literature.

\subsection{Wormhole Attacks}
\label{wormhole-attack}
When an attacker can control two nodes in different VANET segments and has a high speed link between those two, he can mount a so called wormhole attack. Illegal but correct traffic would originate from and to both ends of the tunnel, making vehicles suddenly appear in each others vicinity, while actually being in two remote locations. This type of attack could be the basis for executing other attacks, like sybil \ref{sybil-attack}, eclipse \ref{eclipse-attack} or denial of service \ref{denial-of-service} attacks. A wormhole might be used by an attacker, to generate illegal traffic and let the nodes interfere with each trustworthiness in the connected segments, influence voting procedures or even cause a denial of service when the nodes in both segments revoke each others trustworthiness based on wrong positioning information.

\textbf{Centralized}: Assuming global network visibility is achieved, illegal traffic, which would be generated by a wormhole attack could be spotted by roadside units, acting as a sensor. The central network management system should then be able to correlate that the same traffic is visible in two remote locations. Mitigation of such an attack would only be a notice to affected nodes to discard traffic that is not in their vicinity. Most stronger responses like revoking the right to allocate a channel for communication would not harm the attacker in between but the nodes in their respective network segment. This could result in a DoS attack.

\textbf{Decentralized}: In \cite{Safi2009}, Safi et al. based their effort, like Eichler et al. \cite{Eichler2004}, on the AODV Routing Protocol and enhanced it to include ``geographical leashes'' that should prevent the forwarding of packets from different geographic areas with additional packet authentication.

\textbf{Hybrid}: \textbf{-} No hybrid approaches were found in literature.

\subsection{Denial of Service}
\label{denial-of-service}
Denial of service attacks are often used as distraction or an accompanying attack that should weaken the position of a system to ease the real attack or exploit. This type of issue is one of the harder ones to defend against. Because there are no purely technical means to defend against jamming attacks in wireless communication systems. Types of denial of service attacks include jamming of radio frequencies, traffic flooding or silver bullet attacks, where one specially crafted packet may be able to disrupt service.

\textbf{Centralized}: When a system needs a functioning PKI, like most of the mentioned approaches, or the one from Raya et al. \cite{Raya2007}. A DoS attack could me mounted by creating a lot of identities and then report those same identities as malicious or fraudulent. This could result in a flood of certificate revocations, which could lead to DoS when revocation lists get to big or the revocation operation is computational intensive. To mitigate this threat Raya et al. \cite{Raya2007} suggested the use of ``Compressed Certificate Revocation Lists (RC$^2$RL)'' and ``Revocation of the Trusted Component (RTC)'' protocols.

\textbf{Decentralized}: For VANETs Hamieh et al. \cite{Hamieh2009} described a method to detect on going jamming attacks. They focused on attacks where the jammer is only sending when his hardware is allowed to, he abides the rules of the underlying IEEE 802.11p Standard. Their model is based on time correlation of errors and correct receptions to detect the presence of a jamming attack.

\textbf{Hybrid}: \textbf{-} No hybrid approaches were found in literature.

\subsection{Privacy Violation}
\label{privacy-violation}
In a cooperative system where every neighboring node should have all the needed information to make intelligent decisions on its own and for the group, it is clear that all this information needs to be communicated. Therefore, when every node broadcasts his position, trajectory, acceleration, route or other data, basically a profile of the driver could be created. If this data is readable by everybody in the vicinity, somebody just needs to set up an antenna and can now make statistics where and when people are driving, when traveling past him.
   
\textbf{Centralized}:
To protect privacy and making tracking harder most approaches use pseudonyms and rotating them, like \cite{Sun2007a}, \cite{Choi2009}. But everybody does it slightly different, Sun et al. \cite{Sun2007a} are using ``preloading [...] pseudonym(s)'' whereas Choi et al. \cite{Choi2009} use generation of public keys by deriving it from the secret id only known to an authority and the vehicle itself. While still allowing the verification and certification based on time stamps and other public key parameters. But almost all approaches \cite{Wasef2009a}, \cite{Wasef2009}, \cite{Wasef2010}, found during our survey are using PKI to guarantee authenticity and non-repudiation. The latter one supposedly for liability reasons.

\textbf{Decentralized}: To preserve privacy Li et al. \cite{Li2008} presented a scheme called ``SECSPP'' utilizing non interactive identity based cryptography and a blind signature scheme for allowing anonymous usage of RSU services. Anonymous confidential communication between the RSU and vehicles make tracking or eavesdropping harder and more expensive.

\textbf{Hybrid}: A cluster based architecture utilizing PKI, theshold cryptography and location limited side channel \cite{Balfanz2002}, like license plate recognition is proposed by Bechler et al. \cite{Bechler2004} to secure ad-hoc communication, similar to an approach by Zhou et al. \cite{Zhou1999}. To adapt to different security levels the approach by Bechler et al. \cite{Bechler2004} supports 4 different modes of operation: no encryption, cluster key encryption, public key directly exchanged and public key certified by a distributed certificate authority, in this case the cluster heads.

\section{Conclusion}
\label{conclusion}
Some security issues in ITSs are hard or outright impossible to mitigate on a purely technical basis, this is why we did not consider them in our review. Examples for this type of attacks are, jamming or physical tampering. An attacker with a radio frequency jammer can suppress any meaningful communication \cite{Hubaux2004}. Often the solution to physical tampering is to even better tamper proof those devices, like sensors or Electronic Control Units (ECUs). This climaxes often in the inclusion of a Trusted Platform Module (TPM), which shifts the responsibility and trust to the manufactures of those components. But as explained in the introduction \ref{introduction}, trust in those supposedly highly secure entities has been shattered in the recent years. Therefore relying on them can be the Achilles heal of a system.
Besides those doubts there are many solutions for centralized architectures and some decentralized ones. We were able to find suitable hybrid solutions in the literature for only five out of eight security issues. Often only one hybrid solution could be found for a specific security issue. Our findings are summarized in the Table \ref{table_analysis_overview}. We therefore conclude that hybrid approaches are underrepresented in current research, which might be an indicator that further research is needed or that hybrid approaches appear to be fruitless endeavors. To answer those questions further research, including a comparative study, needs to be conducted. The direction the standardization efforts, by IEEE and ETSI, are heading, is towards centralized architectures with all benefits and weaknesses. Those will set the mark against all other solutions have to prove themselves. Eventually decentralized solutions could be considered for integration in those standards if proven beneficial.  
%\textbf{\color{red}FIXME Weich ja, aber mehr traue mich nicht zu schrieben!}


%\section*{Acknowledgment}
%
%FIXME Acknowledgements do here 



%\bibliographystyle{IEEEtran}
% As suggested below, edit bibtemplate_samples.bib to reflect
% your bibliography. See bibtemplate.text for referencing.
%

\bibliographystyle{IEEEtran}
\bibliography{securware2015}

\end{document}